\documentclass[twoside,12pt]{article}
\usepackage{cite}
\usepackage[centertags]{amsmath}
\allowdisplaybreaks[4]
\usepackage{amsfonts}
\usepackage{amssymb}
\usepackage[dvips]{graphicx}
\usepackage[dvips,hyperindex]{hyperref}
\usepackage{amsmath}

\newcommand{\be}{\begin{equation}}
\newcommand{\ee}{\end{equation}}
\newcommand{\bea}{\begin{eqnarray}}
\newcommand{\eea}{\end{eqnarray}}

\begin{document}
\title{ \vspace{1cm} Matter Effects in Solar Neutrino Active-Sterile Oscillations}
\author{Carlo Giunti$^1$~, Yu-Feng Li$^{1,2}$\footnote{Speaker, li@to.infn.it}\\
$^1$INFN, Sezione di Torino, Via P. Giuria 1, I-10125 Torino, Italy\\
$^2$Department of Modern Physics, University of Science
and\\Technology of China, Hefei, Anhui 230026, China}
\maketitle
\begin{abstract} We study the matter effects for solar neutrino oscillations
in a general scheme, without any constraint on the number of sterile
neutrinos and the mixing matrix elements, only assuming a realistic
hierarchy of neutrino squared-mass differences in which the smallest
squared-mass difference is effective in solar neutrino oscillations.
The validity of the analytic results are illustrated with a
numerical solution of the evolution equation in the simplest case of
four-neutrino mixing with the realistic matter density profile
inside the Sun.
\end{abstract}
Neutrino physics \cite{hep-ph/9812360,Giunti-Kim-2007,0704.1800} is
one of the most active fields in particle physics. The standard
scenario\cite{0704.1800,hep-ph/0405172,hep-ph/0506083} in neutrino
oscillation phenomenology is three neutrino mixing with a
squared-mass hierarchy:
\begin{eqnarray}
\Delta{m}^{2}_{\text{SOL}} \simeq \null & \null 8 \times 10^{-5} \,
\text{eV}^{2} \,,\quad \Delta{m}^{2}_{\text{ATM}} \simeq \null &
\null 2.5 \times 10^{-3} \, \text{eV}^{2} \,. \label{001}
\end{eqnarray}
The existence of a much larger squared-mass difference($\,\gtrsim
0.1 \, \text{eV}^2 \,$), as that indicated by the LSND
\cite{hep-ex/0104049} $\bar\nu_{\mu}\to\bar\nu_{e}$ oscillation
signal, would require the existence of sterile neutrinos in addition
to the three active flavors. However, the LSND signal is currently
disfavored by the negative results of the KARMEN
\cite{hep-ex/0203021}  and MiniBooNE \cite{0812.2243} experiments.
Another indication comes from the anomalous ratio of measured and
predicted ${}^{71}\text{Ge}$ observed in the Gallium radioactive
source experiments GALLEX \cite{Hampel:1998fc} and SAGE
\cite{0901.2200} and the MiniBooNE \cite{0812.2243} low-energy
anomaly, which can be explained by Short Baseline (SBL) electron
neutrino disappearance \cite{0707.4593}. These possible
active-sterile transitions can be tested by studying their effects
in solar neutrino oscillations.\\
The matter effects in solar neutrino active-sterile oscillations
were studied in Ref.\cite{hep-ph/9908513} in a four neutrino scheme,
but the SBL effects were neglected. Thus a combined analysis of the
two anomalies above with the solar neutrino data needs the
derivation of the matter effects in a more generic scheme \cite{0910.5856}.\\
To start, we consider a generic scheme of mixing of three active
neutrino fields ($\nu_{e}$, $\nu_{\mu}$, $\nu_{\tau}$) and $N_{s}$
sterile neutrino fields mixed as $\,\nu_{\alpha L} = \sum_{k=1}^{N}
U_{\alpha k} \nu_{kL}\,(\alpha=e,\mu,\tau,s_{1},\ldots,s_{N_{s}})
\,$, with $\,N=3+N_{s}\,$. Solar neutrinos are described by the
states $\,| \nu(x) \rangle = \sum_{\alpha} \psi_{\alpha}(x) |
\nu_{\alpha} \rangle \,,$ where $x$ is the distance from the
production point with $\,\psi_{\alpha}(0) = \delta_{\alpha e}\,,$
and the amplitudes are normalized as $\,\sum_{\alpha}
|\psi_{\alpha}(x)|^2 = 1\,$. The evolution of the flavor transition
amplitudes $\psi_{\alpha}(x)$ is given by the MSW equation (see
Ref.~\cite{Giunti-Kim-2007}):
\begin{equation}
i \frac{d}{dx} \Psi = \frac{1}{2E} \left( U \mathcal{M}^2
U^{\dagger} + \mathcal{A} \right) \Psi \,, \label{016}
\end{equation}
where $E$ is the neutrino energy and
\begin{align}
\Psi = \null & \null \left(
\psi_{e},\psi_{\mu},\psi_{\tau},\psi_{s_{1}},\ldots,\psi_{s_{N_{s}}}
\right)^T \,, \label{017}
\\
\mathcal{M}^2 = \null & \null
\text{diag}\!\left(0,\Delta{m}^2_{21},\Delta{m}^2_{31},\Delta{m}^2_{41},\ldots,\Delta{m}^2_{N1}\right)
\,, \label{018}
\\
\mathcal{A} = \null & \null
\text{diag}\!\left(A_{\text{CC}}+A_{\text{NC}},A_{\text{NC}},A_{\text{NC}},0,\ldots\right)
\,, \label{019}
\end{align}
with $ \Delta{m}^2_{kj} = m_{k}^2 - m_{j}^2\,$ the squared-mass
differences and $A_{\text{CC}}=2E
V_{\text{CC}} \,, A_{\text{NC}}=2E V_{\text{NC}} \,.$
$\,V_{\text{CC}}$ and $V_{\text{NC}}$ are the charged-current and neutral-current potentials.\\
For solar densities we have
\begin{equation}
A_{\text{CC}} \sim |A_{\text{NC}}| \sim \Delta{m}^2_{21} \ll
|\Delta{m}^2_{k1}| \quad \text{for} \quad k \geq 3 \,. \label{026}
\end{equation}
It is useful to work in the vacuum mass basis $\Psi^{\text{V}} =
\left( \psi^{\text{V}}_{1},\ldots,\psi^{\text{V}}_{N} \right)^T =
U^{\dagger} \Psi \,,$ which satisfies the evolution equation:
\begin{equation}
i \frac{d}{dx} \Psi^{\text{V}} = \frac{1}{2E} \left( \mathcal{M}^2 +
U^{\dagger} \mathcal{A} U \right) \Psi^{\text{V}} \,. \label{028}
\end{equation}
The inequality in Eq.(\ref{026}) imply that the evolution of the
amplitudes $\psi^{\text{V}}_{3},\ldots,\psi^{\text{V}}_{N}$ is
decoupled from the others. Then we have
\begin{equation}
\psi^{\text{V}}_{k}(x) \simeq \psi^{\text{V}}_{k}(0) \, \exp\!\left(
- i \, \frac{ \Delta{m}^2_{k1} x }{ 2 E } \right) \,, \quad
\text{for} \quad k \geq 3 \,, \label{029}
\end{equation}
with $\,\psi^{\text{V}}_{k}(0) = U_{ek}^{*}\,.$ The first two
amplitudes are coupled by the matter effects:
\begin{equation}
i \frac{d}{dx}
\begin{pmatrix}
\psi^{\text{V}}_{1}
\\
\psi^{\text{V}}_{2}
\end{pmatrix}
= \frac{1}{4E}
\begin{pmatrix}
- \Delta{m}^2_{21} + A \cos 2 \xi & A \sin 2 \xi
\\
A \sin 2 \xi & \Delta{m}^2_{21} - A \cos 2 \xi
\end{pmatrix}
\begin{pmatrix}
\psi^{\text{V}}_{1}
\\
\psi^{\text{V}}_{2}
\end{pmatrix}
\,, \label{041}
\end{equation}
with the definitions of $\tan 2 \xi = \text{Y}/\text{X} \,, A =
A_{\text{CC}} \sqrt{ X^2 + Y^2 }\,$, and
\begin{align}
X = \null & \null |U_{e1}|^2 - |U_{e2}|^2 + R_{\text{NC}}
\sum_{\alpha=e,\mu,\tau} \left( |U_{\alpha1}|^2 - |U_{\alpha2}|^2
\right) \,,\quad Y = \null 2 \left| U_{e1}^{*} U_{e2} +
R_{\text{NC}} \sum_{\alpha=e,\mu,\tau} U_{\alpha1}^{*} U_{\alpha2}
\right| \,. \label{037}
\end{align}
From the similarity of (\ref{041}) and the corresponding equation
for $\nu_e$-$\nu_\mu$ or $\nu_e$-$\nu_\tau$ two-neutrino mixing (see
Ref.~\cite{Giunti-Kim-2007}), we obtain the averaged oscillation
probability:
\begin{equation}
\overline{P}_{\nu_{e}\to\nu_{\beta}} = \left[ \frac{1}{2} + \left(
\frac{1}{2} - P_{12} \right) \cos2\vartheta_{\beta}
\cos2\vartheta_{e}^{0} \right] \cos^2\chi_{\beta} \cos^2\chi_{e} +
\sum_{k=3}^{N} |U_{\beta k}|^2 |U_{ek}|^2 \,, \label{068}
\end{equation}
with the mixing angles $\vartheta_{\beta}$ and $\chi_{\beta}$
defined by $|U_{\beta1}|^2 = \cos^2\vartheta_{\beta} \,
\cos^2\chi_{\beta} \,, |U_{\beta2}|^2 = \sin^2\vartheta_{\beta} \,
\cos^2\chi_{\beta} \,,$ and $\,\sin^2\chi_{\beta} = \sum_{k=3}^{N}
|U_{\beta k}|^2 \,.$ The effective mixing angle in the production
region can be written as $\vartheta_{e}^{0} = \vartheta_{e} +
\omega^{0}\,$ with $\tan2\omega^{0} =  (A^{0} \sin 2 \xi) /(
\Delta{m}^2_{21} - A^{0} \cos 2 \xi ) \,.$ The crossing probability
$P_{12}$ is given by
\begin{equation}
P_{12} = \frac { \exp\left( - \frac{\pi}{2} \gamma_{\text{R}} F
\right) - \exp\left( - \frac{\pi}{2} \gamma_{\text{R}}
\frac{F}{\sin^{2}\xi} \right) } { 1 - \exp\left( - \frac{\pi}{2}
\gamma_{\text{R}} \frac{F}{\sin^{2}\xi} \right) } \,
\theta\!\left(A_{0}-A_{\text{R}}\right) \,, \label{071}
\end{equation}
where $\gamma_{\text{R}}$ is the adiabaticity parameter at the
resonance ($A_{\text{R}} = \Delta{m}^2_{21} \cos 2 \xi$):
\begin{equation}
\gamma_{\text{R}} = \frac {\Delta{m}^{2} \sin^{2}2\xi} {2 E \cos2\xi
\left|\text{d}\ln N_{e}/\text{d}x\right|_{\text{R}}} \,. \label{073}
\end{equation}
\begin{figure}[t!]
\begin{center}
\begin{tabular}{cc}
\includegraphics*[bb=11 11 254 208,width=0.40\textwidth]{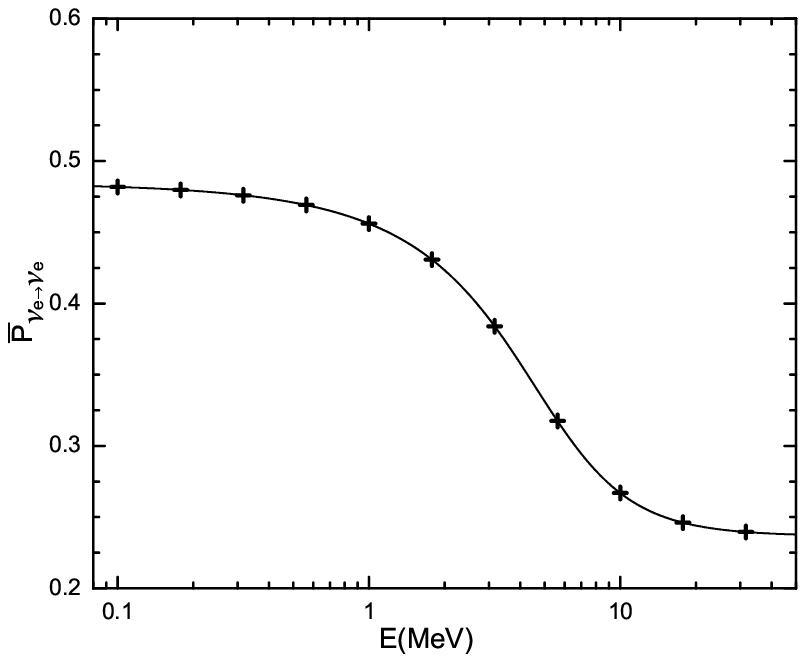}
&
\includegraphics*[bb=18 18 261 215,width=0.40\textwidth]{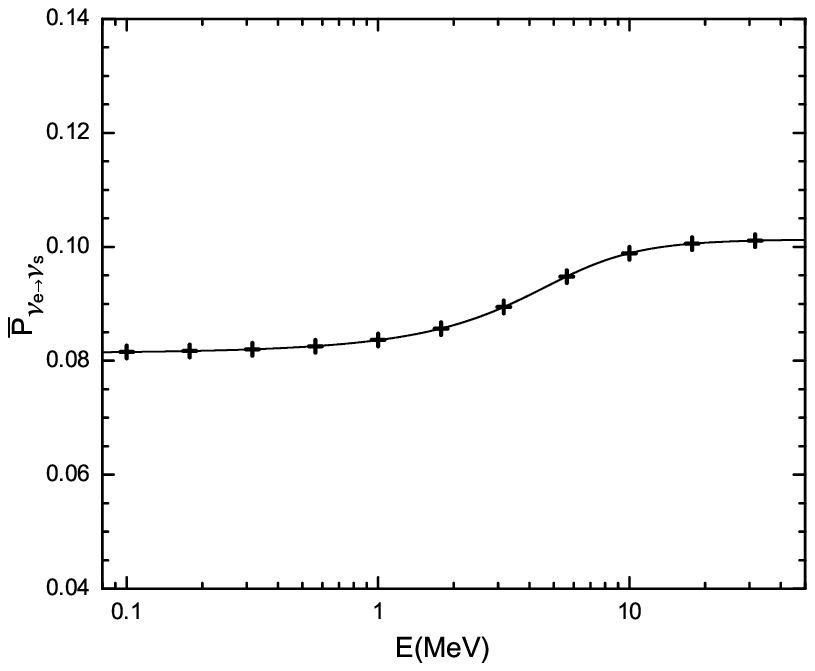}
\end{tabular}
\end{center}
\caption{ \label{fig04} Averaged probability of $\nu_{e}$ survival
and $\nu_{e}\to\nu_{s}$ transitions as functions of the neutrino
energy $E$ for the mixing matrix in (\ref{M1}) calculated for the
BP04 Standard Solar Model density \cite{astro-ph/0402114}. The lines
are obtained with the analytic expression in Eq.~(\ref{068}) and the
overlapping points are obtained with a numerical solution of the
evolution equation. }
\end{figure}
To check the validity of the results for the realistic BP04 Standard
Solar Model density \cite{astro-ph/0402114}, the results of the
analytic expression (\ref{068}) and of the numerical solution of
the evolution equation (\ref{041}) are presented in
Fig.~\ref{fig04}. The relevant mixing parameters are chosen as
$\tan^2 \vartheta_{e} = \tan^2 \vartheta_{\text{SOL}} \simeq 0.4\,$
and
\begin{equation}
\qquad |U_{e3}|^2 = 0.05 \,, \quad |U_{e4}|^2 = 0.05 \,, \quad
|U_{s1}|^2 = 0.03 \,, \quad |U_{s2}|^2 = 0.06 \,. \label{M1}
\end{equation}
We considered only neutrino energies smaller than about 50 MeV, for
which the inequality (\ref{026}) and the approximation (\ref{029})
are valid. In this range of energies, from Fig.~\ref{fig04} one can
see that the analytic approximation is very accurate for solar
neutrinos and can be safely used in the analysis of solar data
\cite{CG-ML-YFL-QYL-2009}.

\end{document}